\documentclass[epj,referee]{svjour}
\usepackage{graphics}
\begin{document}
\title{The relativistic statistical theory and Kaniadakis entropy: an approach through a molecular chaos hypothesis}
\author{R. Silva 
}                     
%
%
\institute{Universidade do Estado do Rio Grande do Norte, 59610-210,
Mossor\'o, RN, Brazil}
\date{Received: date / Revised version: date}
%
\abstract{We have investigated the proof of the $H$ theorem within a
manifestly covariant approach by considering the relativistic
statistical theory developed in [G. Kaniadakis, Phy. Rev. E {\bf
66}, 056125, 2002; {\it ibid.} {\bf 72}, 036108, 2005]. As it
happens in the nonrelativistic limit, the molecular chaos hypothesis
is slightly extended within the Kaniadakis formalism. It is shown
that the collisional equilibrium states (null entropy source term)
are described by a $\kappa$ power law generalization of the
exponential Juttner distribution, e.g., $f(x,p)\propto (\sqrt{1+
\kappa^2\theta^2}+\kappa\theta)^{1/\kappa}\equiv\exp_\kappa\theta$,
with $\theta=\alpha(x)+\beta_\mu p^\mu$, where $\alpha(x)$ is a
scalar, $\beta_\mu$ is a four-vector, and $p^\mu$ is the
four-momentum. As a simple example, we calculate the relativistic
$\kappa$ power law for a dilute charged gas under the action of an
electromagnetic field $F^{\mu\nu}$. All standard results are readly
recovered in the particular limit $\kappa\rightarrow 0$.
} 
\maketitle
\section{Introduction}
\label{intro} The Boltzmann's famous $H$ theorem, which guarantees
positive-definite entropy production outside equilibrium, also
describes the increase in the entropy of an ideal gas in an
irreversible process, by considering the Boltzmann equation. Roughly
speaking, this seminal theorem implies that in the equilibrium
thermodynamic the distribution function of an ideal gas evolves
irreversibly towards maxwellian equilibrium distribution \cite{TO}.
In the special relativistic domain, the very fisrt derivation of
this theorem was done by Marrot \cite{marrot46} and, in the local
form, by Ehlers \cite{ehlers61}, Tauber and Weinberg \cite{tauber61}
and Chernikov \cite{chernikov63}. As well known, the $H$ theorem
furnishes the Juttner distribution function for a relativistic gas
in equilibrium, which contains the number density, the temperature,
and the local four-momentum as free parameters \cite{juttner}.

Recently, this theorem has also been investigated in the context of
a nonextensive statistic mechanics (NSM) \cite{ana92}. In fact, the
NSM has been proposed as a possible extension of the classical one,
being a framework based on the deviations of Boltzmann-Gibbs-Shannon
entropic measure \cite{T95b}. It is worth mentioning that most of
the experimental evidence supporting a NSM are related to the
power-law distribution associated with the many-particle systems
\cite{SPL98}. More recently, based on similar arguments, Abe
\cite{abe97,abe2004} and Kaniadakis \cite{k1,k2} have also proposed
other entropic formulas. In this latter ones, the $\kappa$-entropy
emerges in the context of the special relativity and in the
so-called {\it kinetic interaction principle} (KIP). In particular,
the relativistic $H$ theorem in this approach has also been
investigated through a self-consistent relativistic statistical
theory \cite{kaniadakis05} and through the framework of nonlinear
kinetics \cite{KaniaH01}.

Actually, this $\kappa$-framework leads to a class of one parameter
deformed structures with interesting mathematical properties
\cite{kaniad2001}. In particular, the so-called Lesche stability was
checked in the $\kappa$-framework \cite{kaniadakis04}. It was also
shown that it is possible to obtain a consistent form for the
entropy (linked with a two-parameter deformations of logarithm
function), which generalizes the Tsallis, Abe and Kaniadakis
logarithm behaviours \cite{kania05}. In the experimental viewpoint,
there exist some evidence related with the $\kappa$-statistics,
namely, cosmic rays flux, rain events in meteorology
\cite{kaniad2001}, quark-gluon plasma \cite{miller03}, kinetic
models describing a gas of interacting atoms and photons
\cite{rossani04}, fracture propagation phenomena \cite{cravero04},
and income distribution \cite{drag03}, as well as construct
financial models \cite{bolduc05}. In the theoretical front, some
studies on the canonical quantization of a classical system has also
been investigated \cite{scarfone05}.

From the mathematical viewpoint, the $\kappa$-framework is based on
$\kappa$-exponential and $\kappa$-logarithm functions, which is
defined by
\begin{equation}\label{expk}
\exp_{\kappa}(f)= (\sqrt{1+{\kappa}^2f^2} + {\kappa}f)^{1/{\kappa}},
\end{equation}
\begin{equation}\label{expk1}
\ln_{\kappa}(f)= ({f^{\kappa}-f^{-\kappa})/2\kappa}.
\end{equation}
The $\kappa$-entropy associated with $\kappa$-framework is given by
\begin{equation}\label{e1}
S_{\kappa}(f)=-\int d^{3}p f
[a_{\kappa}f^{\kappa}+a_{-\kappa}f^{-\kappa}+b_{\kappa}],
\end{equation}
which recovers standard Boltzmann-Gibbs entropy
$S_{\kappa=0}(f)=-\int f \ln f d^3 p$ in the limit
$\kappa\rightarrow 0$, i.e, the $S_{\kappa=0}$ is obtained through
the constraints on the constants $a_{\kappa}$ and $b_{\kappa}$ given
by  (see Ref. \cite{k1,k2} for details)
\begin{equation}
\lim_{\kappa\rightarrow 0}[\kappa a_\kappa - \kappa a_{-\kappa}] =1,
\quad \lim_{\kappa\rightarrow 0} [b_\kappa + a_\kappa + a_{-\kappa}]
=0.
\end{equation}
Hereafter the Boltzmann constant is set equal to unity for the sake
of simplicity.

Previous works have already discussed some specific choices for the
constants $a_\kappa$ and $b_\kappa$, i.e., for the pair [$a_\kappa =
1/2\kappa$, $b_\kappa=0$], the Kaniadakis entropy reads
\cite{k1,abe2004}
\begin{equation}\label{first}
S_\kappa = - \int d^3 p f\ln_\kappa f =  - \langle{\ln_\kappa
(f)\rangle}.
\end{equation}
In particular, for entropy (\ref{first}) the $H$ theorem has been
proved in Ref. \cite{k2}. Here, we consider the choice [$a_\kappa=
1/2\kappa(1+\kappa)$, $b_\kappa=-a_\kappa-a_{-\kappa}$] with the
$\kappa$-entropy given by \cite{abe2004}
\begin{equation} \label{tsaent}
S_{\kappa} = - \int{d^{3}p
            \left(\frac{f^{1+\kappa}}{2\kappa(1+\kappa)}
                 -\frac{f^{1-\kappa}}{2\kappa(1-\kappa)}+ b_\kappa f\right)}.
 \end{equation}

In this paper, we intend to extend the nonrelativistic $H$ theorem
within the Kaniadakis framework to the special relativistic domain
through a manifestly covariant approach. As we shall see, our
approach does not consider the so-called deformed mathematics
\cite{k2}. Rather, we show a proof for the $H$ theorem based on
similar arguments of Refs. \cite{silva05,lima01}, e.g., a
generalization of the molecular chaos hypothesis and of the
four-entropy flux.

\section{Classical $H$ Theorem}
\label{sec:1} We first recall the basis for the proof of the
standard $H$ theorem within the special relativity. As well known,
the $H$ theorem is also based on the molecular chaos hypothesis
(Stosszahlansatz), i.e., the assumption that any two colliding
particles are uncorrelated. This means that the two point
correlation function of the colliding particles can be factorized
\begin{equation}\label{Boltzmann2}
 f(x, p, p_1) = f(x, p) f(x, p_1),
\end{equation}
or, equivalently,
\begin{equation}\label{Boltzmann2a}
 \ln f(x, p, p_1) = \ln f(x, p) + \ln f(x, p_1),
\end{equation}
where $p$ and $p_1$ are the four-momenta just before collision and
the particles have four-momentum $p\equiv p^\mu=(E/c,\bf {p})$ in
each point $x \equiv x^\mu=(c t,\bf{r})$ of the space-time, with
their energy satisfying $E/c=\sqrt{{\bf p}^2+m^2 c^2}$. In order to
complete the proof of the $H$ theorem, we combine the Boltzmann
equation with the four-divergence of the four-entropy flow, i.e.,
\begin{equation}
S^{\mu}=-c^2 \int {d^{3}p\over E} p^{\mu} f \ln f.
\end{equation}

In this concern, it is possible to show that the relativistic
Kaniadakis entropy is consistent with a slight departing from
``Stosszahlansatz". Basically, this means the replacement of the
logarithm functions appearing in (\ref{Boltzmann2a}) by
$\kappa$-logarithmic (power laws) defined by Eq. (\ref{expk1}). In
reality, it is worth mentioning that the validity of the chaos
molecular hypothesis still remains as a very controversial issue
\cite{Zeh}.

\subsection{Generalized $H$ Theorem}
\label{sec:2} In order to investigate $H$ theorem in the context of
the Kaniadakis statistics, we first consider a relativistic rarified
gas containing $N$ point particles of mass $m$ enclosed in a volume
$V$, under the action of an external four-force field $F^\mu$.
Naturally, the states of the gas must be characterized by a Lorentz
invariant one-particle distribution function $f(x,p)$, which the
quantity $f(x,p) d^3xd^3p$ gives, at each time $t$, the number of
particles in the volume element $d^3xd^3p$ around the particles
space-time position $x$ and momentum ${\bf p}$. By considering that
every influence of a power law statistic must happen within the
collisional term of Boltzmann equation (see also
\cite{silva05,lima01}), we assume that the temporal evolution of the
relativistic distribution function $f(x,p)$ is given by the
following $\kappa$-transport equation
\begin{equation}\label{relBot}
p^{\mu}\partial_\mu f+m F^\mu{\partial{f}\over\partial
p^{\mu}}=C_{\kappa}(f),
\end{equation}
where $\mu = 0,1,2,3$ and $\partial_\mu=(c^{-1}\partial_t,\nabla)$
indicates differentiation with respect to time-space coordinates and
$C_\kappa$ denotes the relativistic $\kappa$-collisional term.
Following  the same physical arguments concerning the collisional
term from approach of Refs. \cite{silva05,lima01}, we have that
$C_\kappa(f)$ has the general form
\begin{equation}
C_\kappa(f) = {c\over 2} \int F \sigma R_\kappa (f,f'){d^3 p_1\over
E_1} d\Omega,
\end{equation}
where $d\Omega$ is an element of the solid angle, the scalar $F$ is
the invariant flux, which is equal to $F=\sqrt{(p_\mu p^\mu_1)^{2}
-m^4 c^4}$, and $\sigma$ is the differential cross section of the
collision $p + p_1\rightarrow p' + p'_1$; see Ref. \cite{degrrot}
for details. All quantities are defined in the center-of-mass system
of the colliding particles. Next, we observe that $C_\kappa$ must be
consistent with the energy, momentum, and the particle number
conservation laws, and its specific structure must be such that the
standard result is recovered in the limit $\kappa\rightarrow 0$.

Here,  the $\kappa$-generalized form of molecular chaos hypothesis
is also a difference of two correlation functions
\begin{eqnarray*} \label{eq:2.12}
R_{\kappa} (f,f') = \exp_{\kappa}\left(\ln_{\kappa}{ f'}+
                          \ln_{\kappa}{
                          f'_{1}}\right)
\end{eqnarray*}
\begin{equation} -
            \exp_{\kappa}\left(\ln_{\kappa}{ f} +
                          \ln_{\kappa}{f_{1}}\right)
,
\end{equation}
where primes refer to the distribution function after collision, and
$\exp_\kappa (f)$, $\ln_\kappa(f)$, are defined by Eqs. (1) and (2).
Note that for $\kappa= 0$, the above expression reduces to
$R_0=f'{f'}_1-ff_1$, which is exactly the standard molecular chaos
hypothesis. In the present framework, the $\kappa$ four-entropy flux
reads
\begin{equation}
S^{\mu}_\kappa = - c^2\int{{d^{3}p\over E}
            p^\mu\left(\frac{f^{1+\kappa}}{2\kappa(1+\kappa)}
                 -\frac{f^{1-\kappa}}{2\kappa(1-\kappa)} +  b_\kappa f\right)},
\end{equation}
where for $\mu = 0$, the quantity $c^{-1}S^0_\kappa$ stands for the
local Kaniadakis` entropy density, as given by (\ref{tsaent}).
Indeed, by taking the four-divergence of $S^\mu_\kappa$, i.e.,
\begin{equation}
\partial_\mu S^{\mu}_\kappa \equiv \tau_\kappa= -c^{2} \int{d^3p\over E} p^\mu\partial_\mu f(\ln_\kappa f+b_\kappa),
\end{equation}
and combining with $\kappa$ relativistic Boltzmann equation
(\ref{relBot}), we obtain the source term
\begin{equation}
\tau_\kappa = -{c^3\over 2} \int{d^3p\over E}{d^3p_1\over
E_1}d\Omega F \sigma R_\kappa(\ln_\kappa f + b_\kappa).
\end{equation}
Next, $\tau_\kappa$ can be written in a more symmetrical form by
using some elementary symmetry operations, which also take into
account the inverse collisions. Let us notice that by interchanging
$p$ and $p_1$ the value of the integral above is not modified. This
happens because the scattering cross section and the magnitude of
the flux are invariants \cite{degrrot}. The value of $\tau_\kappa$
is not altered if we integrate with respect to the variables $p'$
and $p'_1$. Although changing the sign of $R_\kappa$ in this step
(inverse collision), the quantity ${d^3pd^3p_1/p^0 p^0_1}$ is also a
collisional invariant \cite{degrrot}. As one may check, such
considerations imply that the $\kappa$-entropy source term can be
written as
\begin{eqnarray*}  \label{eq: 2.26}
\tau_{\kappa}(x)={c^3\over 8} \int {d^3p\over E}{d^3p_1\over
E_1}d\Omega F \sigma (\ln_{\kappa}{f'_{1}} +\ln_{\kappa}{f'}-
\end{eqnarray*}
\begin{eqnarray*}
-\ln_{\kappa}{f_{1}} -\ln_{\kappa}{f})[\exp_{\kappa}(\ln_{\kappa}{
f'}+\ln_{\kappa}{f'_{1}})-
\end{eqnarray*}
\begin{eqnarray}
-\exp_{\kappa}(\ln_{\kappa}{f} + \ln_{\kappa}{f_{1}})].
\end{eqnarray}

As is well known, the irreversibility thermodynamics emerging from
molecular collisions is quite obtained if $\tau_\kappa(x)$ is
positive definite. This condition is guaranteed only when the
integrand in (\ref{eq: 2.26}) is not negative. Indeed, by
introducing the auxiliary functions, namely
$z=\exp_{\kappa}(\ln_{\kappa}f+\ln_{\kappa}{f}_1)$ and
$y(z)=\ln_{\kappa}f+\ln_{\kappa} {f}_1.$ The differences $z'-z$ and
$y'-y$ can be positive or negative and these differences have the
same sign if the functions $y$ is an increasing function. However,
$\ln_\kappa f$ is an increasing function and then product
$(y'-y)(z'-z)$ is always positive for any pair of distributions
$(f,f_1)$ and $(f',f'_1)$. Therefore, the positiveness of
$\tau_\kappa$ must be established  \cite{degrrot,GP}.

For the sake of completeness, let us derive the version of the
Juttner distribution within the $\kappa$-statistic. Such an
expression is the relativistic version of the $\kappa$-distribution
\cite{kaniad2001}, and must be obtained as a natural consequence of
the relativistic $H$ theorem. At this point, it is interesting to
emphasize that such a distribution already has been introduced by
Kaniadakis through a variational problem in a selfconsistent
approach; see Ref. \cite{kaniadakis05} for details. The $H$ theorem
states that $\tau_\kappa=0$ is a necessary and sufficient condition
for equilibrium. Since the integrand of (\ref{eq: 2.26}) cannot be
negative, this occurs if and only if
\begin{equation}
\ln_{\kappa} f'+\ln_{\kappa} {f}'_1=\ln_{\kappa}f+\ln_{\kappa} f_1,
\end{equation}
where the four-momenta are connected through a conservation law
$$p^\mu + p^\mu_1 = {p'}^\mu+{p'}^\mu_1,$$ which is valid for any
binary collision. Therefore, the above sum of $\kappa$-logarithms
remains constant during a collision: it is a summational invariant.
In the relativistic case, the most general collisional invariant is
a linear combination of a constant plus the four-momentum $p^\mu$;
see Ref. \cite{degrrot}. Consequently, we must have
\begin{equation}\label{joia}
\ln_{\kappa} f (x,p)=\theta=\alpha(x)+\beta_\mu p^\mu,
\end{equation}
where $\alpha(x)$ is a scalar, $\beta_\mu$ a four-vector, and
$p^\mu$ is the four-momentum. By using that $\exp_\kappa (\ln_\kappa
f) = f$, we may rewrite (\ref{joia}) as
\begin{equation} \label{eq:2.34a}
f(x,p)=\exp_{\kappa}\theta=(\sqrt{1+
\kappa^2\theta^2}+\kappa\theta)^{1/\kappa},
\end{equation}
with arbitrary space and time-dependent parameters $\alpha(x)$ and
$\beta_\mu(x)$. Some considerations on the function $f(x,p)$ are
given as follows. First, this is the most general expression which
leads to a vanishing collision term and entropy production, and
reduces to Juttner distribution in the limit ${\kappa \rightarrow 0}
$. However, it is not true in general that $f(x,p)$ is a solution of
the transport equation. This happens only if $f(x,p)$ also makes the
left-hand-side of the transport equation (\ref{relBot}) to be
identically null. Nevertheless, since (\ref{eq:2.34a}) is a power
law, the transport equation implies that the parameters $\alpha(x)$
and $\beta_\mu (x)$ must only satisfy the constraint equation
\begin{equation}
p^\mu\partial_\mu \alpha(x)+p^\mu p^\nu\partial_\mu
\beta_\nu(x)+m\beta_\mu(x) F^\mu(x,p)=0.
\end{equation}
Second, the above expression is also the relativistic version of the
Kaniadakis distribution \cite{k1}. Here, this was obtained through
the different approach from the one used in Refs.
\cite{k2,kaniadakis05}.

As an example, let us now  consider a relativistic gas under the
action of the Lorentz 4-force $$F^\mu(x,p)=-(q/mc)
F^{\mu\nu}(x)p_\nu,$$ where $q$ is the charge of the particles and
$F^{\mu\nu}$ is the Maxwell electromagnetic tensor. Following
standard lines, it is easy to show that the local equilibrium
function in the presence of an external electromagnetic field reads
\begin{equation}
f(x,p)=\exp_{\kappa}\left[{\mu-[p^\mu+c^{-1}q A^\mu (x)]U_\mu\over
T}\right],
\end{equation}
where $U_\mu$ is the mean four-velocity of the gas, $T(x)$ is the
temperature field, $\mu$ is the Gibbs function per particles, and
$A^\mu(x)$ the four potential. As physically expected, note also
that the above expression reduces, in the limit $\kappa \rightarrow
0$, to the well known expression \cite{degrrot,Hakim}
\begin{equation}
f(x,p)=\exp\left({\mu-[p^\mu+c^{-1}q A^\mu (x)]U_\mu\over T}\right).
\end{equation}

\section{Conclusions}
In Refs. \cite{silva05,lima01} we have discussed the $H$ theorem in
the context from the Kaniadakis and Tsallis statistics within the
nonrelativistic and relativistic domain. Based on the generalization
of the chaos molecular hypothesis and the entropic measure, it was
shown the proof of the $H$ theorem in both domain. In this paper, by
considering the same arguments on the chaos molecular and entropy,
and regardless of the deformed mathematics introduced in Ref.
\cite{k2}, we have studied a $\kappa$-generalization of the
relativistic Boltzmann's kinetic equation along the lines defined by
the Kaniadakis statistics. In reality, since the basic results were
obtained through a manifestly covariant way, their generalization to
the general relativistic domain may be readly derived. Finally, we
also emphasize that our proof is consistent with the standard laws
describing the microscopic dynamics, and reduce to the familiar
Boltzmann proof in the limit $\kappa\rightarrow0$.

This work was supported by Conselho Nacional de
Desenvolvimento Cient\'{i}fico e Tecnol\'{o}gico - CNPq (Brasil).

%
%
%
%
 \bibliography{}
%

\end{document}